# Ro-vibrational relaxation of HCN in collisions with He:
# Rigid bender treatment of the bending-rotation interaction

Thierry Stoecklin,[1*] Otoniel Denis-Alpizar,[1,2] Philippe Halvick,[1] and Marie-Lise Dubernet[3,4]

[1)] *Université de Bordeaux, ISM, CNRS UMR 5255, 33405 Talence Cedex, France*
[2)] *Departamento de Física, Universidad de Matanzas, Matanzas 40100, Cuba*
[3)] *Université Pierre et Marie Curie, LPMAA, UMR CNRS 7092, 75252 Paris, France*
[4)] *Observatoire de Paris, LUTH, UMR CNRS 8102, 92195 Meudon, France*

**Abstract**

We present a new theoretical method to treat atom-rigid bender inelastic collisions at the Close Coupling level (RBCC) in the space fixed frame. The coupling between rotation and bending is treated exactly within the rigid bender approximation and we obtain the cross section for the rotational transition between levels belonging to different bending levels. The results of this approach are compared with those obtained when using the rigid bender averaged approximation (RBAA) introduced in our previous work dedicated to this system. We discuss the validity of this approximation and of the previous studies based on rigid linear HCN.

Corresponding author: t.stoecklin@ism.u-bordeaux1.fr

# 1. INTRODUCTION

Because of their importance to model the chemistry of interstellar clouds, quantum inelastic scattering calculations involving small polyatomic molecules are the subject of many theoretical studies [1]. However most of them are limited to the use of the rigid rotor approximation as it is expected to be a quite accurate approach to calculate rotational transitions which are the most probable at the typical temperature of interstellar clouds. This is the case of most available studies of energy transfer collisions involving an atom and a linear [2] or bent triatomic molecule [3]. The neglect of the vibration motion relies on the fact that the vibrational frequencies are commonly quite large compared to the rotational ones. The validity of this approximation was recently confirmed at low collision energy by the excellent agreement between experiment and calculation obtained for the $H_2$-CO inelastic cross sections [4]. However, several authors pointed out that infrared transition among the molecular vibrational levels could significantly increase the intensities of the rotational transitions by populating the upper rotational levels [5]. The question of the validity of this approximation was then recently investigated by the group of Alexander and Dagdigian for $CH_2$ [6]. They calculated the bending levels of $CH_2$ and averaged the He-$CH_2$ potential over the bending angle using these functions in order to reduce the problem to a fictitious atom-rigid asymmetric top molecule collision. In a recent paper [7], hereafter denoted paper I, we used a similar approach which we called Rigid Bender Average Approximation (RBAA) for a collision involving this time He and a linear molecule: HCN. This last molecule and isocyanide (HNC) are among the most abundant organic molecules in the interstellar medium. The averaging over the bending angle of the intermolecular potential reduced the calculation in our case to even simpler atom linear molecule calculations. The bending frequency of HCN is only the half of the one of $CH_2$ suggesting that the exact treatment of the coupling between rotation and bending may be more important for this molecule. Therefore, the present study is dedicated to the development of a method including exactly the coupling between bending and rotation for a collision between an atom and a rigid bender. This study was furthermore motivated by recent astrophysical measurements of vibrationaly excited HCN [8,9] in the interstellar medium. The authors assumed that the molecule is pumped to the excited bending level by infrared radiation and return to the vibrational ground state with rotational excitation [10,5]. What makes the exact treatment even more necessary is the recent detection of HCN *l*-type transitions in hot planetary nebula [11] as it involves nearly degenerate levels. Such transitions can be calculated using the RBAA approach but it seems important to check if the results given by this approximation are reliable by comparing with the exact results.

In section 2 of the present paper we give the close coupling equations for atom rigid bender collisions. In section 3, we compare our results to those given by the RBAA approach and for pure rotational transitions with atom-linear HCN close coupling results.

## 2. METHOD

We need first to calculate the bending levels of HCN alone for each value of its rotational angular momentum. While the problem of calculating the rovibrational states of a triatomic molecule in internal coordinates was solved long ago and can be performed using the MORBID code of Jensen [12] we decided to give here a short overview of the corresponding equations which we coded as the literature of that time is full of contradicting terms and misprints.

### A. Calculation of the HCN rigid bender energies and wave functions in internal coordinates

We solve the variational problem using the rigid bender Hamiltonian of Sutcliffe [13] which for the Z molecular axis is along the intermolecular coordinate between H and the center of mass of CN:

$$H = \left[ K_V^{(1)} + K_V^{(2)} + K_{VR} \right] + V(\gamma) \qquad (1)$$

Where $K_V^{(1)}$ and $K_V^{(2)}$ are the contributions to the kinetic operator which are independent of rotation while $K_{VR}$ is the rotation-bending interaction term. The bending angle is denoted as $\gamma$ and $V(\gamma)$ is the 1-dimensional bending potential of HCN reported in paper I

$$K_V^{(1)} = \left[ -\frac{\hbar^2}{2} \left( \frac{1}{\mu_1 R_1^2} + \frac{1}{\mu_2 R_2^2} \right) \left( \left( \frac{\partial^2}{\partial \gamma^2} \right) + \cot\gamma \frac{\partial}{\partial\gamma} \right) \right] \qquad (2)$$

$$K_V^{(2)} = \frac{\hbar^2}{\mu_{12} R_1 R_2} \left[ \cos\gamma \left( \left( \frac{\partial^2}{\partial\gamma^2} \right) + \cot\gamma \frac{\partial}{\partial\gamma} \right) + \sin\gamma \frac{\partial}{\partial\gamma} \right] \qquad (3)$$

$$K_{VR} = \frac{1}{2} \left[ \left( \frac{1}{\mu_1 R_1^2} \right) \cot^2\gamma + \left( \frac{1}{\mu_2 R_2^2} \right) \frac{1}{\sin^2\gamma} - \frac{2\cos\gamma}{\sin^2\gamma} \left( \frac{1}{\mu_{12} R_1 R_2} \right) \right] \hat{\Pi}_Z^2$$
$$+ \left( \frac{1}{2\mu_1 R_1^2} \right) \left[ \hat{\Pi}_X^2 + \hat{\Pi}_Y^2 \right] + \left[ \left( \frac{1}{2\mu_1 R_1^2} \right) \frac{\cos\gamma}{\sin\gamma} - \left( \frac{1}{2\mu_{12} R_1 R_2} \right) \frac{1}{\sin\gamma} \right] \left[ \hat{\Pi}_X \hat{\Pi}_Z + \hat{\Pi}_Z \hat{\Pi}_X \right] \qquad (4)$$
$$+ \frac{\hbar}{i} \left\{ \left( \frac{1}{\mu_1 R_1^2} \right) \left[ \frac{\partial}{\partial\gamma} + \frac{1}{2}\cot\gamma \right] - \left( \frac{1}{2\mu_{12} R_1 R_2} \right) \left[ \frac{1}{\sin\gamma} + 2\cos\gamma \frac{\partial}{\partial\gamma} \right] \right\} \hat{\Pi}_Y$$

with $\frac{1}{\mu_1} = \frac{1}{M_C} + \frac{1}{M_N}$ ; $\frac{1}{\mu_2} = \frac{1}{M_C} + \frac{1}{M_H}$ ; $\mu_{12} = M_C$

and where $\hat{\Pi}_x, \hat{\Pi}_y, \hat{\Pi}_z$ are the projection over the molecule fixed axis of the rotational angular momentum of the HCN molecule.

We then follow Sutcliffe and Tennyson [14] and first take the matrix elements of $K_{VR}$ in a symmetric top basis set $|jKM\rangle = \sqrt{\frac{(2J+1)}{8\pi^2}} D_{M,K}^{j*}(\phi_{R_{HCN}}, \theta_{R_{HCN}}, 0)$ where $M$ and $K$ are the projections of the rotational angular momentum $j$ of HCN along the Z space fixed axis and along the Z molecular fixed axis respectively, and where $(\phi_{R_{HCN}}, \theta_{R_{HCN}})$ are the spherical coordinates of the vector $\vec{R}_{HCN}$ joining H to the center of mass of CN in the space fixed frame:

$$\langle jK'M | K_{VR} | jKM \rangle = \delta_{K,K'} \left[ \left( \frac{1}{2\mu_1 R_1^2} \right) \left[ J(J+1) - 2K^2 \right] + \frac{K^2}{2} \frac{1}{\sin^2\gamma} \left[ \frac{1}{\mu_1 R_1^2} + \frac{1}{\mu_2 R_2^2} - \frac{2\cos\gamma}{\mu_{12} R_1 R_2} \right] \right]$$

$$+ \delta_{K+1,K'} C_{jK}^+ \left( \frac{1}{2\mu_1 R_1^2} \right) \left[ -\frac{\partial}{\partial\gamma} + K\cot\gamma \right] + \delta_{K-1,K'} C_{jK}^- \left( \frac{1}{2\mu_1 R_1^2} \right) \left[ \frac{\partial}{\partial\gamma} + K\cot\gamma \right]$$

$$+ \delta_{K+1,K'} C_{jK}^+ \left( \frac{1}{2\mu_{12} R_1 R_2} \right) \left[ \cos\gamma \left( \frac{\partial}{\partial\gamma} - K\cot\gamma \right) - K\sin\gamma \right] \quad (5)$$

$$+ \delta_{K-1,K'} C_{jK}^- \left( \frac{1}{2\mu_{12} R_1 R_2} \right) \left[ -\cos\gamma \left( \frac{\partial}{\partial\gamma} + K\cot\gamma \right) - K\sin\gamma \right]$$

with $C_{JK}^\pm = \sqrt{J(J+1) - K(K\pm1)}$

The second term of this expression cancels some other terms appearing in $K_V^{(1)} + K_V^{(2)}$. The terms remaining in $K_{VR}$ are splitted into the two following expressions:

$$\langle jK'M | K_{VR}^{(1)} | jKM \rangle = \delta_{K,K'} \left[ \left( \frac{\hbar^2}{2\mu_1 R_1^2} \right) \left[ J(J+1) - 2K^2 \right] \right]$$

$$+ \delta_{K+1,K'} C_{jK}^+ \left( \frac{\hbar^2}{2\mu_1 R_1^2} \right) \left[ -\frac{\partial}{\partial\gamma} + K\cot\gamma \right] + \delta_{K-1,K'} C_{jK}^- \left( \frac{\hbar^2}{2\mu_1 R_1^2} \right) \left[ \frac{\partial}{\partial\gamma} + K\cot\gamma \right] \quad (6)$$

and

$$\langle jK'M | K_{VR}^{(2)} | jKM \rangle = \left( \frac{\hbar^2}{2\mu_{12} R_1 R_2} \right) \left[ \delta_{K+1,K'} C_{jK}^+ \left[ \cos\gamma \left( \frac{\partial}{\partial\gamma} - K\cot\gamma \right) - K\sin\gamma \right] \right.$$

$$\left. + \delta_{K-1,K'} C_{jK}^- \left[ -\cos\gamma \left( \frac{\partial}{\partial\gamma} + K\cot\gamma \right) + K\sin\gamma \right] \right] \quad (7)$$

We now take the matrix elements of these terms in a normalized associated legendre polynomial basis set $\tilde{P}_l^k(\cos\gamma)$ as defined by Green [15].

$$Y_l^m(\gamma,\varphi) = (-)^m \tilde{P}_l^k(\cos\gamma) e^{im\varphi}$$

and obtain for the different terms:

$$\left\langle \tilde{P}_{l'}^{K'}(\cos\gamma) jK'M \middle| K_V^{(1)} \middle| \tilde{P}_l^K(\cos\gamma) jKM \right\rangle = \frac{\hbar^2}{2} \delta_{K,K'} \delta_{l,l'} \left[ \frac{1}{\mu_1 R_1^2} + \frac{1}{\mu_2 R_2^2} \right] \quad (8)$$

$$\left\langle \tilde{P}_{l'}^{K'}(\cos\gamma) jK'M \middle| K_V^{(2)} \middle| \tilde{P}_l^K(\cos\gamma) jKM \right\rangle = \frac{-\hbar^2}{\mu_{12} R_1 R_2} \delta_{K,K'} \delta_{l',l\pm1} \sqrt{\frac{(l_> + m)(l_> - m)}{(2l_> + 1)(2l_> - 1)}} \left\{ \begin{array}{c} l^2 \\ (l+1)^2 \end{array} \right\} \quad (9)$$

In this last expression the top and the bottom terms enclosed in square brackets are respectively associated with the values $l'=l+1$ and $l'=l-1$ while $l_>$ is the maximum of $l$ and $l'$.

$$\left\langle \tilde{P}_{l'}^{K'}(\cos\gamma) jK'M \middle| K_{VR}^{(1)} \middle| \tilde{P}_l^K(\cos\gamma) jKM \right\rangle = \frac{\hbar^2}{2\mu_1 R_1} \delta_{l',l} \left[ \delta_{K,K'} \left[ J(J+1) - 2K^2 \right] \right.$$
$$\left. - \delta_{K',K+1} C_{JK}^+ C_{lK}^+ - \delta_{K',K-1} C_{JK}^- C_{lK}^- \right] \quad (10)$$

$$\left\langle \tilde{P}_{l'}^{K'}(\cos\gamma) jK'M \middle| K_{VR}^{(2)} \middle| \tilde{P}_l^K(\cos\gamma) jKM \right\rangle = \frac{\hbar^2}{2\mu_{12} R_1 R_2} \left[ \delta_{K',K+1} C_{jK}^+ \left[ C_{lK}^+ \delta_{l',l\pm1} \sqrt{\frac{(l_> + K + 1)(l_> - K - 1)}{(2l_> + 1)(2l_> - 1)}} \right. \right.$$
$$\left. - K \left[ a_{lK} \delta_{l',l+1} - b_{lK} \delta_{l',l-1} \right] \right] + \delta_{K',K-1} C_{jK}^- \left[ C_{lK}^- \delta_{l',l\pm1} \sqrt{\frac{(l_> - K + 1)(l_> + K - 1)}{(2l_> + 1)(2l_> - 1)}} + K \left[ a_{l-K} \delta_{l',l+1} - b_{l-K} \delta_{l',l-1} \right] \right] \right]$$

(11)

where

$$a_{lK} = \sqrt{\frac{(l + K + 1)(l + K + 2)}{(2l + 3)(2l + 1)}} \quad \text{and} \quad b_{lK} = \sqrt{\frac{(l - K - 1)(l - K)}{(2l - 1)(2l + 1)}}$$

while the potential is expended in Legendre polynomials :

$$V(\gamma) = \sum_p C_p P_p(\cos\gamma)$$

$$\left\langle \tilde{P}_{l'}^{K'}(\cos\gamma) jK'M \middle| V(\gamma) \middle| \tilde{P}_l^K(\cos\gamma) jKM \right\rangle =$$
$$\delta_{K',K} \sum_p C_p (-)^K \sqrt{(2l+1)(2l'+1)} \begin{pmatrix} l' & p & l \\ 0 & 0 & 0 \end{pmatrix} \begin{pmatrix} l' & p & l \\ -K & 0 & K \end{pmatrix} \quad (12)$$

For each value of $j$, the diagonalisation of the resulting hamiltonian matrix gives the rigid bender energies $\varepsilon_{vj}$ and the corresponding space fixed rigid bender eigenfunctions

$$\chi_v^{jM}(\gamma) = \sum_K \sum_{l \geq K} C_{l,K}^{vj} \tilde{P}_l^K(\cos\gamma) \middle| jKM \right\rangle = \sum_K \Gamma_{j,K}^v(\gamma) \middle| jKM \right\rangle \quad (13)$$

where $v$ designates the bending quantum number. This equation shows that the rovibrational functions obtained from the rigid bender model can be put in the form of a product of an asymmetric top rotational function by a vibrational bending function $\Gamma_{j,K}^v(\gamma)$. This is in agreement with the intuitive representation that the bending vibration of a linear triatomic molecules makes it become an instantaneous asymmetric top, the square of the $T^{\nu}_{j,K}(\gamma)$ function giving the probability of a given bending angle for a specific values of the quantum numbers associated with the bending state $\nu$ the rotational state $j$ and its projection along the the molecular axis K.

## B. Close Coupling equations

The theory of the inelastic scattering of two rigid polyatomic molecules was developed long ago [16] but studies including both vibration and rotation of the fragments are seldom. Most of such studies use the pioneering approaches developed by Clary [17] or more recently by Bowman[18] which rely on the use of the Infinite order sudden approximation for the rotation and the Close Coupling for the vibration (VCC-IOS). We treat here the scattering of a rigid bender molecule colliding with an atom. This approach takes only into account the coupling between bending and rotation but could be easily generalized to include the stretching vibrations. We use the result of the previous paragraph to express the rovibrational wave function of the He-HCN complex in space fixed coordinates as:

$$|\nu\, jlJM\rangle = \sum_K \Gamma^{\nu}_{j,K}(\gamma) \sum_{m_j} \sum_{m_l} \sqrt{\frac{2J+1}{4\pi}} \langle jm_j lm_l | JM \rangle D^{j*}_{m_j,K}(\phi_{R_{HCN}}, \theta_{R_{HCN}}, 0) Y^{m_l}_l(\hat{R}) \quad (14)$$

where $\hat{R}$ are the spherical coordinates of the intermolecular vector $\vec{R}$ between He and the center of mass of HCN in the space fixed frame. Following Green [19] we now expand the intermolecular potential between the atom and the rigid bender molecule in the body fixed coordinates defined in Fig.1 of paper I like:

$$V_{A-RB}(R,\gamma,\theta,\varphi) = \sum_\lambda \sum_\mu v_{\lambda\mu}(R,\gamma) Y^{\mu}_{\lambda}(\theta,\varphi) \quad (15)$$

which can be written:

$$V_{A-RB}(R,\gamma,\theta,\varphi) = \sum_\lambda \sum_{\mu\geq 0} v_{\lambda\mu}(R,\gamma)\left[2-\delta_{\mu 0}\right]\tilde{P}^{\mu}_{\lambda}(\cos\theta)\cos(\mu\varphi) \quad (16)$$

Expression (11) transformed by rotation in terms of space fixed angles reads:

$$V_{A-RB}(R,\gamma,\phi_{R_{HCN}},\theta_{R_{HCN}},\hat{R}) = \sum_\lambda \sum_\mu v_{\lambda\mu}(R,\gamma) \sum_\nu D^{\lambda*}_{\mu\nu}(\phi_{R_{HCN}},\theta_{R_{HCN}},0) Y^{\mu}_{\lambda}(\hat{R}) \quad (17)$$

The matrix elements of the interaction potential $V_{A-RB}$ between the atom and the rigid bender molecule in the space fixed coordinates are obtained from straightforward algebra:

$$[V_{A-RB}]^{JM}_{vjl,v'j'l'}(R) = \langle vjlJM | V_{A-RB} | v'j'l'JM \rangle =$$
$$\sum_k \sum_{k'} \sum_\lambda \sum_\mu \int_0^\pi d\gamma \sin\gamma \Gamma^v_{j,k}(\gamma) V_{\lambda\mu}(R,\gamma) \Gamma^{v'}_{j',k'}(\gamma) W^{JM}_{jkl,j'k'l'} \quad (18)$$

where the $W^{JM}_{jkl,j'k'l'}$ are the atom symmetric top matrix elements in space fixed coordinates [19]

$$W^{JM}_{jkl,j'k'l'} = \delta_{J,J}\delta_{M,M} \cdot (-1)^{[j+j'+\lambda+J]} \left[(2j+1)(2j'+1)(2l+1)(2l'+1)(2\lambda+1)\right]^{\frac{1}{2}}$$
$$\sum_\lambda \sum_\mu V_{\lambda\mu}(R,\gamma) \begin{Bmatrix} l & J & j \\ J' & l' & 1 \end{Bmatrix} \begin{pmatrix} l & \lambda & l' \\ 0 & 0 & 0 \end{pmatrix} \begin{pmatrix} j & \lambda & j' \\ -k & -\mu & k' \end{pmatrix} \quad (19)$$

The Close Coupling equations to solve for an atom-triatomic rigid bender collision and for given values of $J$ and $M$ are then:

$$\left[\frac{d^2}{dR^2} - \frac{l(l+1)}{R^2} + k^2_{vj}(E) - [V_{A-RB}]^{JM}_{vjl,v'j'l'}(R)\right] G^{JM}_{vjl,v'j'l'}(R)\Big|_I = 0 \quad (20)$$

where $k^2_{vj}(E) = 2\mu_{He-HCN}[E - \varepsilon_{nj}]$ and $\varepsilon_{nj}$ are the rigid bender energies calculated previously. From this expression we see that the atom-triatom rigid bender equations are similar in form to those of the atom linear molecule system, the only differences being the matrix elements of the potential and the rigid bender energies. Indeed if we put $k=k'=0$ and fix $\gamma=180°$ we obtain the usual atom-linear molecule Close Coupling equations in space fixed coordinates [20]. In this expression we did not introduced the parity of HCN as it is broken by the bending vibration. This last point shows that the results of the RBAA approach which uses the parity of linear HCN cannot be completely equivalent to those given by the RB-CC method for transitions between different bending levels as discussed below.

## 3. CALCULATIONS AND RESULTS

In the following we compare the results obtained with the atom-linear molecule close coupling (ALM-CC), RBAA and RB-CC approaches and using our potential [7]. In all kind of calculations we use 20 rotational levels of HCN. For the RB-CC and RBAA calculations we include in the basis set the three first bending levels for each rotational level of HCN. For each value of K (the projection of the rotational angular momentum of HCN along the Z molecular fixed axis), we use 30 associated Legendre functions to calculate the bending levels using the equations presented in section 2. The He-HCN dynamics is performed using a Gauss legendre quadrature of 40 values of the bending angle.

We first consider the pure rotational transitions taking place inside the fundamental bending level $v=0$. We represented in Figures 1, 2 and 3 the elastic and de-excitation cross sections respectively of the levels ($v=0$, $j=1, 2, 3$) of HCN. For each transition represented, we can observe a group of two curves which are almost identical and are associated with the ALM-CC and RBAA approaches while the third curve associated with the present RB-CC results is slightly different. The differences between the cross sections obtained from the three types of calculations are in any case negligible in the [1:1000] cm$^{-1}$ energy interval needed for Astrochemistry. Only the very low energy range and the regions of the resonances are slightly different. This demonstrate that for these transitions the ALM-CC approach offers a level of accuracy equivalent to the RBAA approach, as already concluded in paper I, and is, for the He-HCN system, almost equivalent to exact calculations.

As mentioned in the introduction, $l$-type transitions of HCN have been detected recently in hot molecular gas, for example in the proto-planetary nebula CRL 618. The double degeneracy of the bending mode of HCN is indeed lifted when the molecule is bending and rotating simultaneously giving ride to $l$-type doubling for $j\geq1$. For the first excited bending mode and for $j\geq1$, every rotational level is splitted into two sub-levels. Hereinafter we designate these two sub-levels by $v=1$ and $v=2$ and examine the $l$-type transitions between these sub-levels. In Figs. 4 and 5, the elastic and the de-excitation cross sections starting from the ($v=2$, $j=1$) level of HCN respectively calculated using the RBAA and RB-CC approaches are represented as a function of the collision energy. The upper panels of each of these two figures show all these transitions while the lower panels show a blow up including only the transitions which are not attributed in the upper panel. The only open rotational channel of de-excitation towards the bending level $v=1$ is $j=1$ as the ($v=2$, $j=1$) and ($v=1$, $j=1$) levels are almost degenerate while the first excited bending energy of the level $j=0$, which is exempt from $l$-type doubling, is quite higher. The open rotational channel of de-excitation towards the bending level $v=0$ are the levels $j=0$-19, $j=19$ being the highest value of $j$ considered in our calculations. As it can be seen in these figures, the RBAA gives at the best the right order of magnitude for some of the transitions but is definitely not accurate enough even for astrochemical purposes. The elastic and $l$-type transition cross sections (towards the levels ($v=2$, $j=1$) and ($v=1$, $j=1$)) are for example found to be almost equal at the RBAA level while they differ by an order of magnitude at the RB-CC level. As the number of transitions represented makes the comparison difficult we compare in Fig. 6 the two types of results for a few selected transitions. We can see on this figure that the RBAA approach fails to give an accurate estimate of the magnitude of the cross sections but gives at least the right ranking of the transitions cross sections. Not surprisingly, we find with both methods that the $\Delta j=0$ transitions are favored and that the magnitude of the cross section decreases when $\Delta j$ increases. Another interesting feature can be seen when comparing Figs. 4 and 5. A dip in the RBAA cross sections appear around 700 cm$^{-1}$ on many curves while it is absent from the RB-CC curves. This can be understood when looking at Figs. 7 and 8 where the state selected ($v=2, j=1 \rightarrow v=0, j=17$) transition cross sections are represented respectively at the RBAA and RB-CC levels. Clearly the dip appears at the RBAA level while it is absent of the RB-CC cross section. On the same figure the cross section associated with the opening of the ($v=1, j=17$) and ($v=2, j=17$) channels are represented. These two levels are strongly coupled with the channel ($v=0, j=17$) as they both are linked by the $\Delta j=0$ rule. We can see in Fig. 7 that the minimum of the dip of the RBAA curve corresponds precisely to the opening of the ($v=0, j=17$) channel while the RB-CC ($v=2, j=1 \rightarrow v=0, j=17$) cross section (Fig. 8) is reduced but shows no dip. We conclude from these different comparisons that the RBAA approach overestimates the coupling between the bending levels. This analysis if confirmed when comparing the RBAA and RB-CC ($v=2, j=1 \rightarrow v=0, j=1$) transition cross sections which are also represented in Figs. 7 and 8. The RB-CC cross sections are effectively lower than those calculated using the RBAA approach. This is not really surprising as we use parity selected matrix elements of the potential in the RBAA approach whereas this definition of the parity is valid only for collisions involving rigid linear molecules.

## 4. CONCLUSION

We presented a new method for calculating exactly ro-vibrational cross sections for collisions between an atom and a rigid bender triatomic molecule. This approach was applied to the He-HCN collision and its results were compared to those obtained when using the rigid bender averaged approximation. We find for this system that the RBAA approach is almost equivalent to exact calculations for pure rotational transitions taking place inside the fundamental bending level $v=0$. On the contrary, for transitions involving two different bending levels, the RBAA approach fails to give an accurate estimate of the magnitude of the cross sections but gives most of the time at least the right ranking of the transition cross sections. We then conclude that $l$-type transitions cross sections have to be calculated at the RB-CC level for the He-HCN collision while pure rotational transitions cross sections may be calculated accurately at the RBAA level. This result should hold for other triatomic molecules which have a similar bending frequency but needs to be tested on other systems.

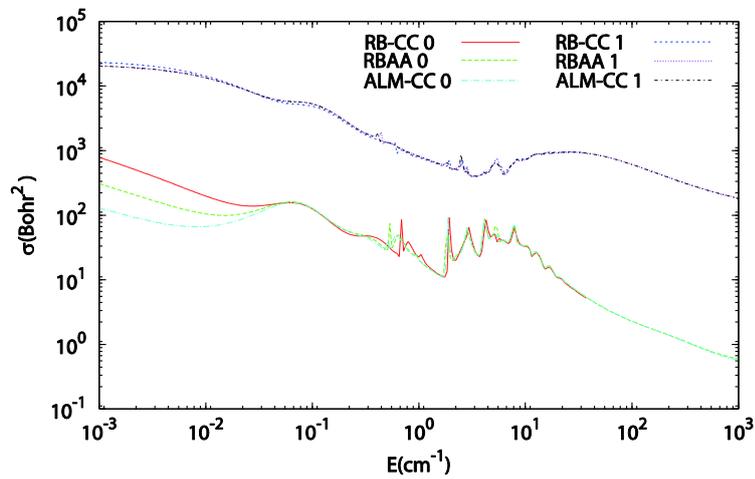

**Figure 1**: Comparison of the elastic de-excitation cross section of HCN($v$=0, $j$=1) in collisions with He as a function of collision energy calculated using the RB-CC, RBAA and ALM-CC approaches. The final level is indicated by one integer designating the rotational quantum number.

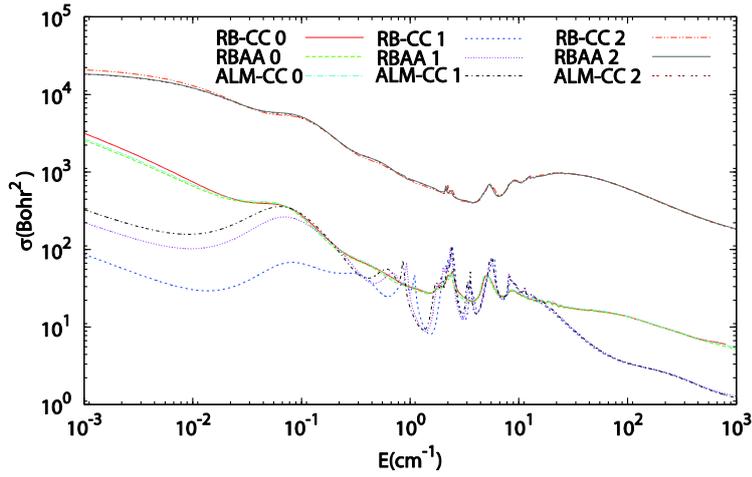

**Figure 2**: Comparison of the elastic and de-excitation cross section of HCN($v$=0, $j$=2) in collisions with He as a function of collision energy calculated using the RB-CC, RBAA and ALM-CC approaches. The final level is indicated by one integer designating the rotational quantum number.

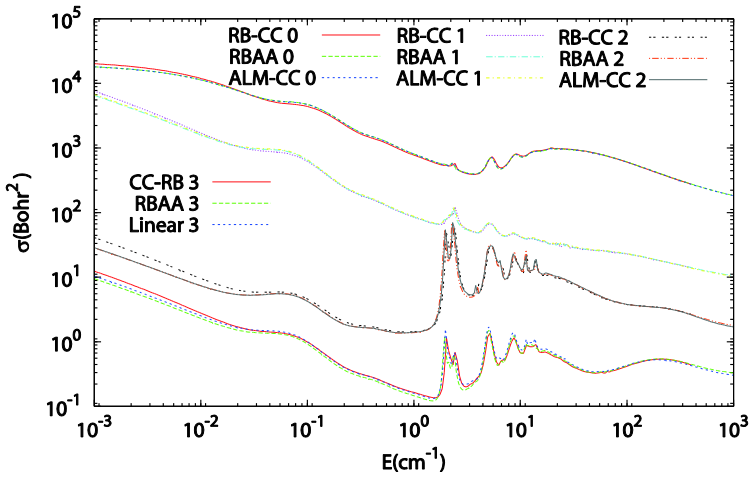

**Figure 3**: Comparison of the elastic and de-excitation cross section of HCN($v$=0, $j$=3) in collisions with He as a function of collision energy calculated using the RB-CC, RBAA and ALM-CC approaches. The final level is indicated by one integer designating the rotational quantum number.

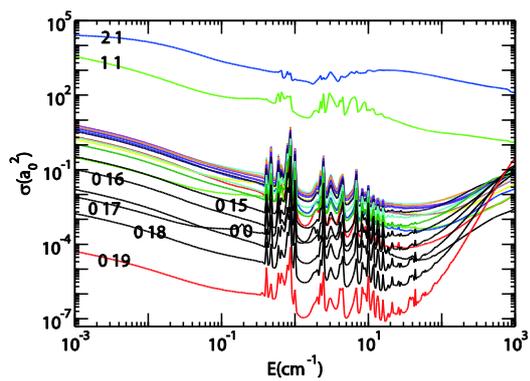
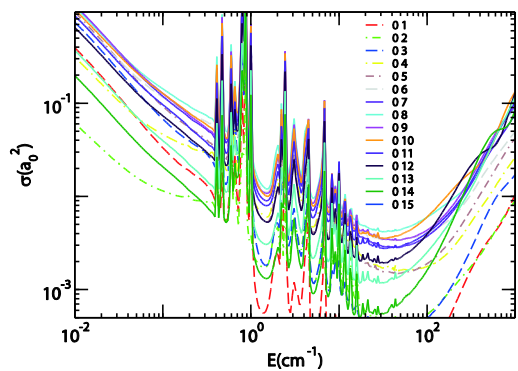

**Figure 4**: Elastic and de-excitation RB-CC cross section of HCN($v$=2, $j$=1) in collisions with He as a function of collision energy. The final level is indicated by two integers designating the bending and the rotational quantum numbers.

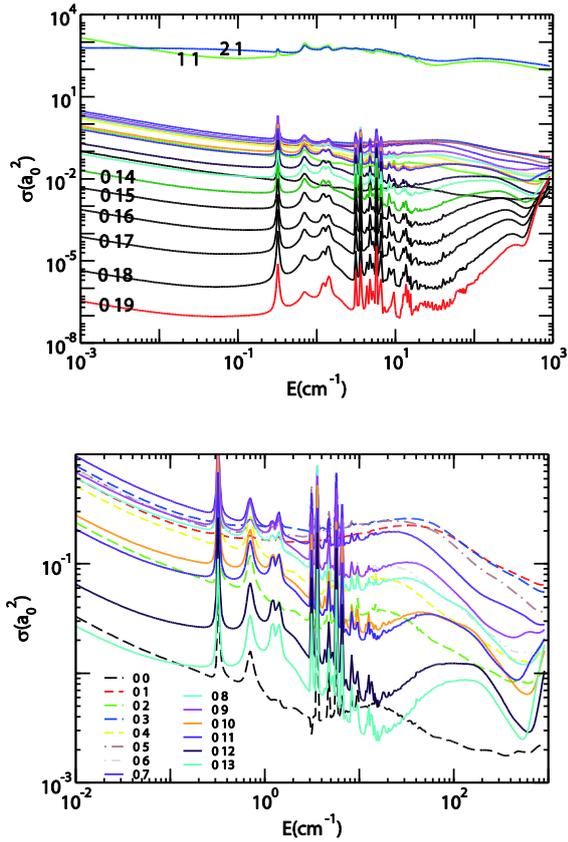

**Figure 5**: Elastic and de-excitation RBAA cross section of HCN($v$=2, $j$=1) in collisions with He as a function of collision energy. The final level is indicated by two integers designating the bending and the rotational quantum numbers.

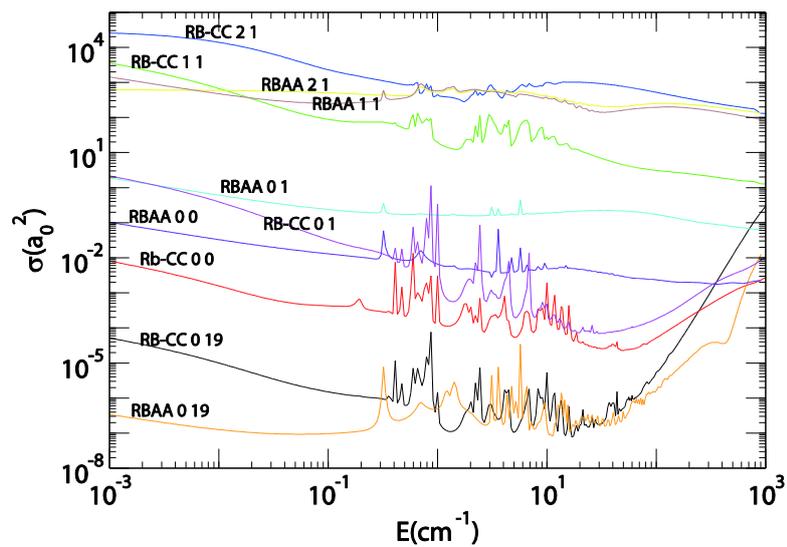

**Figure 6**: Comparison of some of the de-excitation RBAA and RB-CC cross section of HCN($v$=2, $j$=1) in collisions with He as a function of collision energy. The final level is indicated by two integers designating the bending and the rotational quantum numbers.

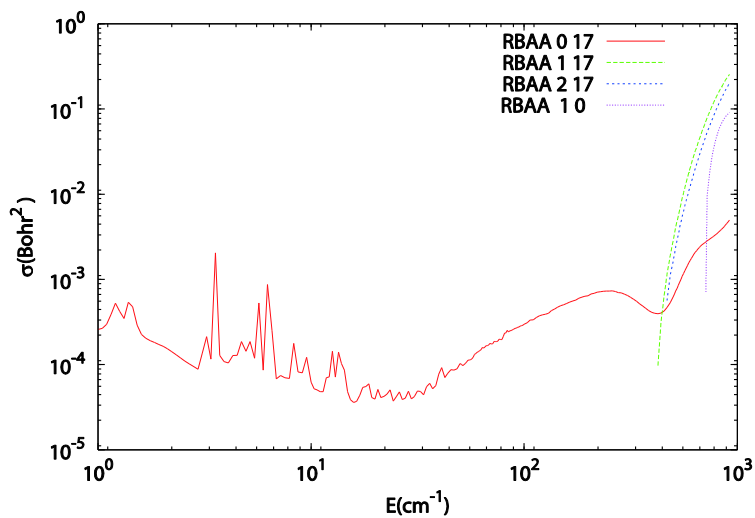

**Figure 7**: Comparison of the inelastic RBAA cross section of HCN($v=2$, $j=1 \rightarrow v'$, $j'=17$) in collisions with He as a function of collision energy. The final level is indicated by two integers designating the bending and the rotational quantum numbers. The cross section associated with the opening of the first excited bending level of the rotational state $j'=0$ of HCN is also represented.

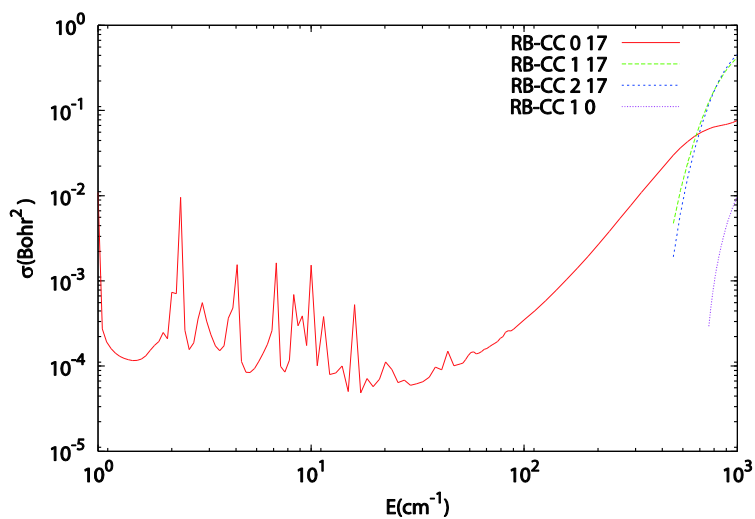

**Figure 8**: Comparison of the inelastic RB-CC cross section of HCN($v=2$, $j=1 \rightarrow v'$, $j'=17$) in collisions with He as a function of collision energy. The final level is indicated by two integers designating the bending and the rotational quantum numbers. The cross section associated with the opening of the first excited bending level of the rotational state $j'=0$ of HCN is also represented.